\documentclass[a4paper]{article}
 
\usepackage{graphicx}
 
\usepackage{amsmath}
 
\usepackage{hyperref}
 
\usepackage{color}

\begin{document}

\begin{center}

{\bf \Large The Simmel effect and babies names}\\[5mm]

{\large M. J. Krawczyk, A. Dydejczyk and K. Ku{\l}akowski}\\[3mm]

{\em
 
 Faculty of Physics and Applied Computer Science, AGH University of
 Science and Technology, 
al. Mickiewicza 30, PL-30059 Krak\'ow,
 Poland\\

 }


{\tt kulakowski@fis.agh.edu.pl}

\bigskip

\today

\end{center}

\begin{abstract}

Simulations of the Simmel effect are performed for agents in a scale-free social network. 
The social hierarchy of an agent is determined by the degree of her node. Particular features, once selected 
by a highly connected agent, became common in lower class but soon fall out of fashion and extinct.  
Numerical results reflect the dynamics of frequency of American babies names in 1880-2011.

\end{abstract}

\noindent

{\em PACS numbers:} 89.65.Ef;  07.05.Tp

\noindent

{\em Keywords:} fashion, simulation, babies names, Simmel effect

\bigskip

\section{Introduction}

Fashion as an object of research was introduced to sociology at the beginning of XX century 
by Georg Simmel \cite{sim}, in direct connection with social classes. According to Simmel, fashion "is a product 
of class distinction".  Dynamics of fashion is driven by two forces (we would prefer to say "processes"): adaptation
to society and individual departure from its demands. The social stratification is projected into a division of
roles: elites tend to differ from lower classes, while the latter tend to imitate elites. These processes of
imitation and avoidance produce a stream of given status symbols (clothing, social conduct, amusement) from elites to 
lower classes, where finally they disappear, replaced by new patterns. Later, the phenomenon was called "Simmel 
effect" \cite{con1,con2}.\\

While imitation as an object of social simulations has attracted common interest \cite{cmv,sznajd,deff,galam,santo}, 
the thread of avoidance is much less popular.  The original Simmel thoughts were converted to simulations by Roberto Pedone
and Rosaria Conte \cite{con1,con2}, but these works remain almost unnoticed. In the original version of the famous model 
of dissemination of culture by Robert Axelrod \cite{axe,cmv} avoidance is absent, and it has been added only 
recently \cite{radi,mjk}. Similarly, social repulsion appeared to be a useful concept when added \cite{huet} 
to the Deffuant model of dynamics of social opinion \cite{deff}. \\

It is somewhat surprising that the authors of these papers do not compare their results with real data on fashion. 
The conclusions of \cite{con1,con2} are concentrated on the fact that the Simmel effect is present in the numerical 
results, and on the mutual comparison of different variants of calculations. Current interpretations of the results of 
the Axelrod model seem to follow large scale theory, as suggested by the term 'cultural' \cite{dybiec}. In \cite{cmv},
the authors suggest that their numerical results could be related to the distribution of languages. However,
in their model the number of equivalent options of cultural traits (the variable $q$) is of the order of hundreds. 
Obviously, nobody in this world has a choice of one hundred languages. The interpretation of fashion (as for example 
clothing), although natural here, remains unexplored.\\

The aim of this text is to connect the calculations of Pedone and Conte to some sets of real data on fashion, available 
in literature. Namely, we intend to apply the model to the datasets of American babies names in the period 1880-2010 
\cite{names}. Some comments will also be possible on the data on skirt lengths \cite{curran,barbie,low}. The model itself 
is slightly modified; a scale-free network is used as the structure of a model social network, instead of a square lattice 
\cite{con1} or a torus \cite{con2}. The number of options $q$ remains as a model parameter, as in \cite{cmv}; however, we 
do not take into account any interaction between different variables, so the number of variables ($F$ in \cite{cmv,radi}) 
is set to one. The social status of agents remains constant during the simulation and is read from the node degree.\\

In two subsequent sections, the model is explained and numerical results are shown. Section 4 is devoted to the datasets,
and Section 5 - to the Simmel effect in the data on babies names. In the last section, we summarize the similarities and 
differences between the results and the data. 

\section{The model}

Scale-free networks are constructed with new nodes attached to $M$ nodes, according to the known principle of preferential 
attachment \cite{preat}. The network size was kept large enough to assure the mean shortest path not less than 3; for example, for $M=5$
the minimal network size is $M=1000$ nodes. As noted above, the social status of nodes is determined by their degree.
The network structure should be complex enough to contain nodes of middle class between elite nodes and nodes of low class;
hence the distances in networks should be large enough.\\

A variable $x_i$ is assigned to each node. The values of these variables belong to the set $1,2,...,q$. The parameter $q$
marks then the number of options of a cultural feature. In the initial state $x_i=1$ for all nodes. \\

For each node $i$, its neighbours are divided into two sets: those with degree larger or equal to the degree of $i$ form
the set $i+$, and the remaining neighbours -- the set $i-$. A node $k$ which has no neighbours in $k+$ is marked. During 
the simulation, nodes are selected randomly. For the selected node, say $j$, states of the set $j+$ are estimable, and states 
of the set $j-$ are shaming. If a state (a number $y$ within the allowed range $1,...,q$) exists which is simultaneously 
estimable and not shaming, it is legal. Then we substitute $x_j=y$. For marked nodes, all states are estimable. If more 
legal states exist, $x_j$ takes the value which is most frequent in $j+$. If no legal states exist for $j$, $x_j$ remain 
unchanged. For marked nodes, the legal state is chosen randomly. To speed up the simulation, we start from one of marked 
nodes.\\ 

This part of algorithm is taken from \cite{con2}, where the authors write: "The symbols of higher level neighbors will be
marked as legal only if they are not equal to the symbols of the lower level neighbors. The agent will modify its color if 
(a) no symbol of higher level neighbors is equal to its own or (b) one at least of the lower level neighbors is equal to 
it. In either case, the agent will randomly assume one of the remaining legal colors." The difference between our 
algorithm and the algorithm in \cite{con2} is that we use the scale-free network and not the torus. Also, in our approach
high degree is equivalent to high social status, while in \cite{con2} it is assigned randomly. This means that in 
\cite{con2}, the structure of interacting nodes is not coupled with the status; in our approach, this coupling is present.
Indications, that this coupling exists, abound in the literature; for a review see \cite{borg}.

\section{The results}

A set of scale-free networks is investigated, of $N$=1000 and 5000, $M$=3, 5 and 8. For all networks, the number of options
$q=30$ appears to be sufficiently large to observe the same behaviour as for all higher values of $q$. Namely, once a new 
value $x$ appears at a marked node, the number of nodes described with this value increases, later it decreases. In most 
observed cases, finally it decreases to zero. This does not prevent a reappearance of this value in subsequent cycles. The 
cyclic behaviour, rise and fall of some values, is visualised in Fig. 1. \\

 \begin{figure}[ht]
 \centering
 {\centering \resizebox*{12cm}{9cm}{\rotatebox{-90}{\includegraphics{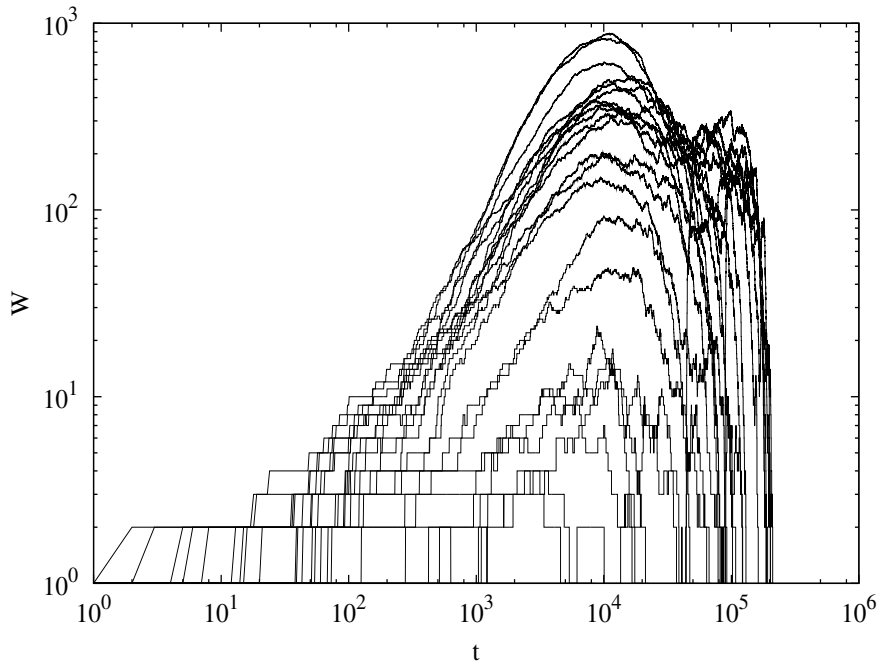}}}}
\caption{A bunch of popularities $W$ of given symbols against time $t$ for $q$=30 - results of the simulation. On the vertical axis we show the number of nodes described with a given symbol. For each curve, the time where it starts is shifted to one.}
 \label{fig-1}
 \end{figure}

 \begin{figure}[ht]
 \centering
 {\centering \resizebox*{12cm}{9cm}{\rotatebox{-90}{\includegraphics{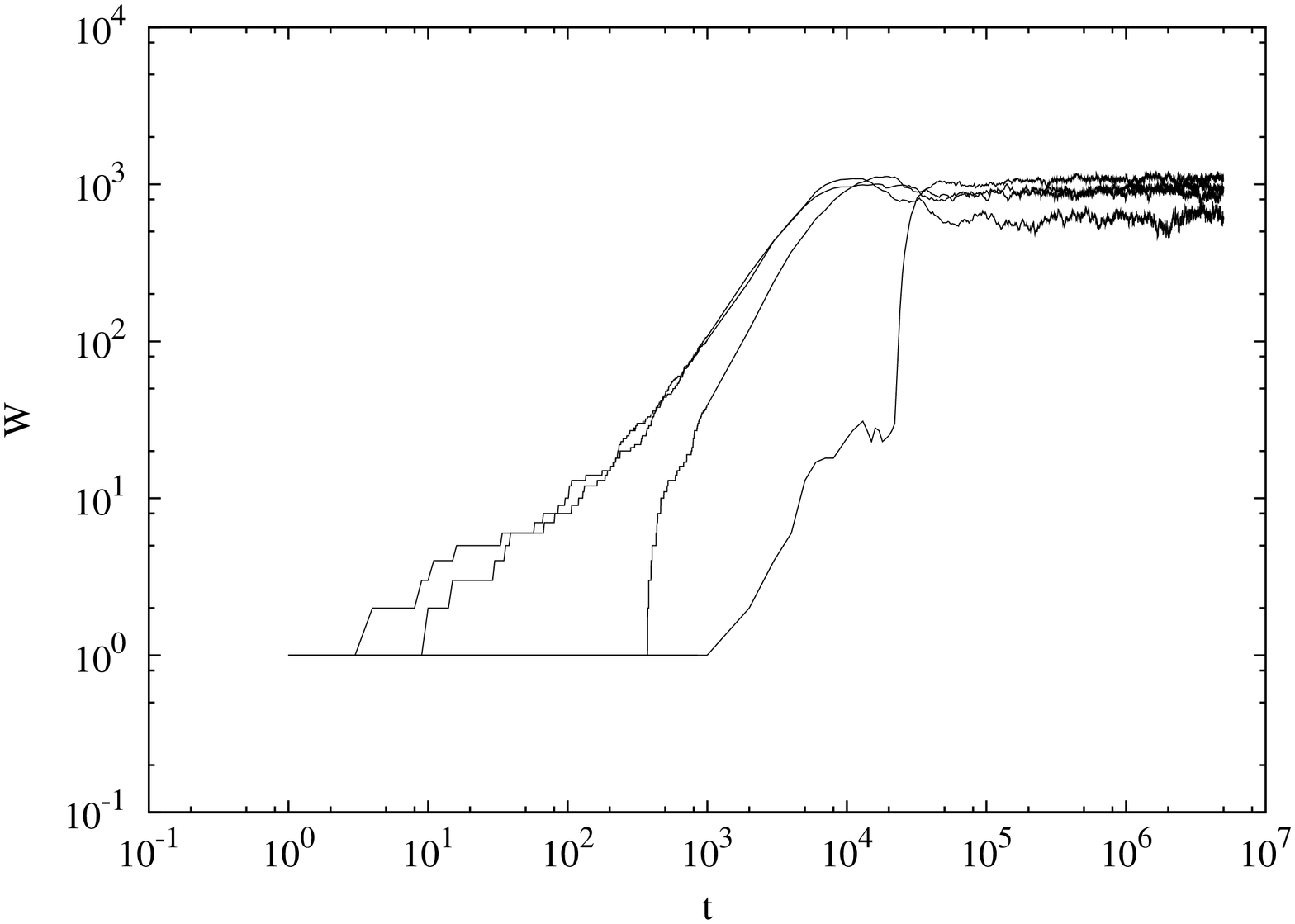}}}}
\caption{A bunch of popularities $W$ of given symbols against time $t$ for $q$=5 - results of the simulation. On the vertical axis we show the number of nodes described with a given symbol. For each curve, the time where it starts is shifted to one.}
 \label{fig-2}
 \end{figure}

 \begin{figure}[ht]
 \centering
 {\centering \resizebox*{12cm}{9cm}{\rotatebox{-90}{\includegraphics{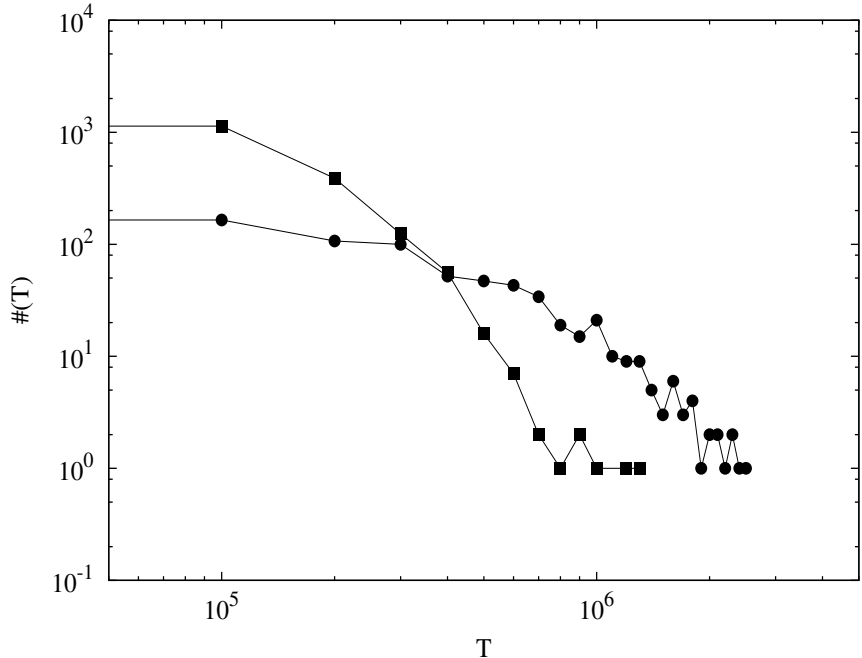}}}}
\caption{Calculated histograms $\#(T)$ of time $T$ of duration of cycles for $q$=30 (circles) and $q$=1000 (squares). }
 \label{fig-7}
 \end{figure}

 \begin{figure}[ht]
 \centering
 {\centering \resizebox*{12cm}{9cm}{\rotatebox{-90}{\includegraphics{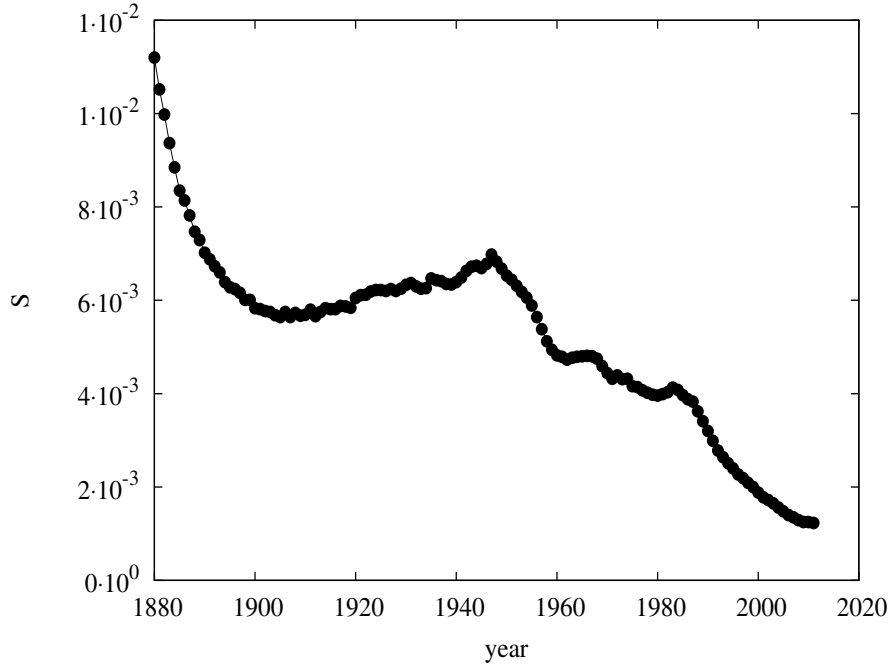}}}}
\caption{Time dependence of the fragmentation $S$ for babies names in U.S. in 1880-2011 \cite{names}. The plot shows a systematic decrease except a small maximum at the end of WWII.}
 \label{fig-3}
 \end{figure}

 \begin{figure}[ht]
 \centering
 {\centering \resizebox*{12cm}{9cm}{\rotatebox{-90}{\includegraphics{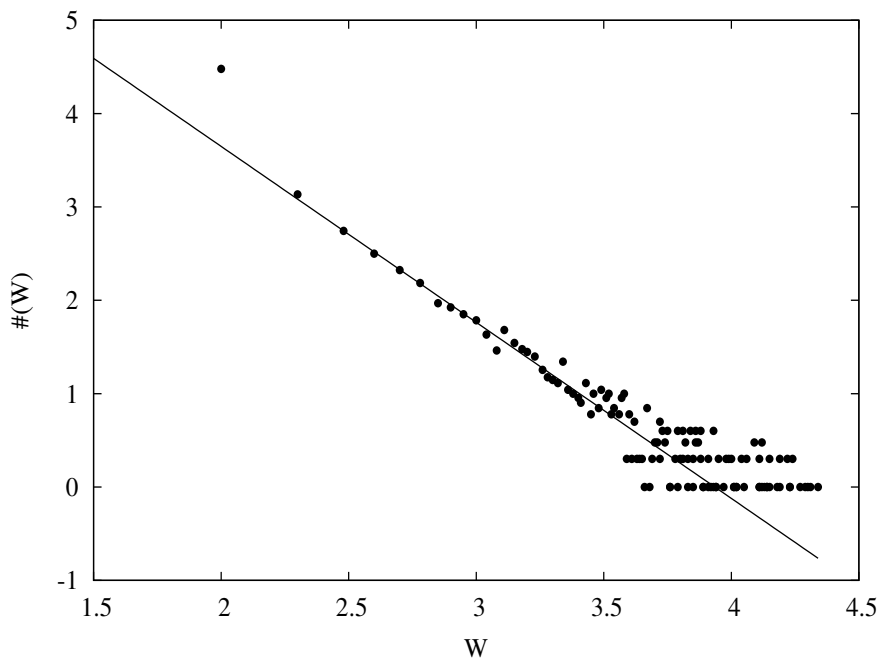}}}}
\caption{The histogram $\#(W)$ of number $W$ of newborn babies with given name in U.S. in 2011 \cite{names}. The plot is close to a power law, except the number of very rare names, which is even larger.}
 \label{fig-4}
 \end{figure}

 \begin{figure}[ht]
 \centering
 {\centering \resizebox*{12cm}{9cm}{\rotatebox{-90}{\includegraphics{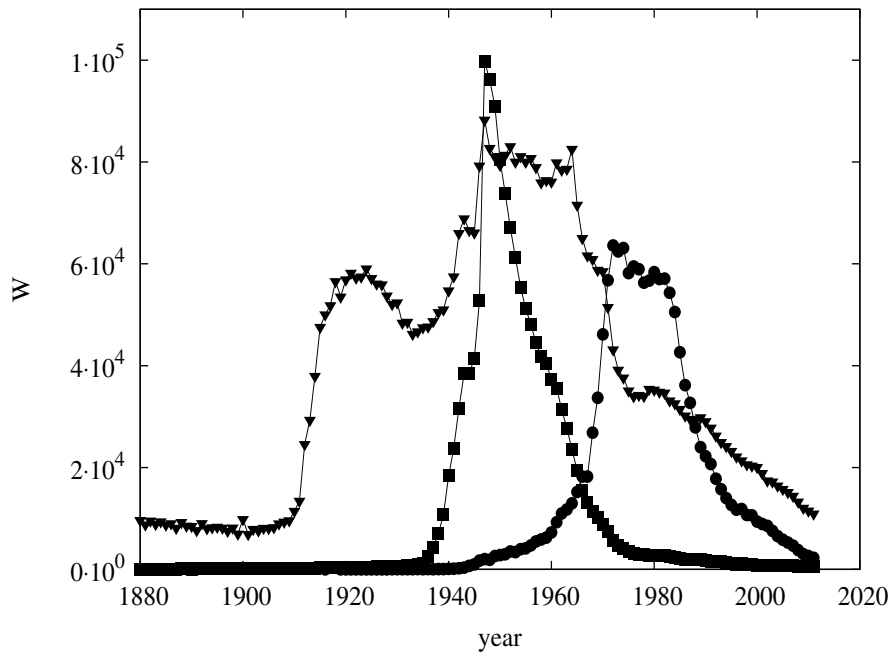}}}}
\caption{Year dependence of number $W$ of newborn children named John (triangles), Linda (squares) and Michael (circles) \cite{names}. }
 \label{fig-5}
 \end{figure}

 \begin{figure}[ht]
 \centering
 {\centering \resizebox*{12cm}{9cm}{\rotatebox{-90}{\includegraphics{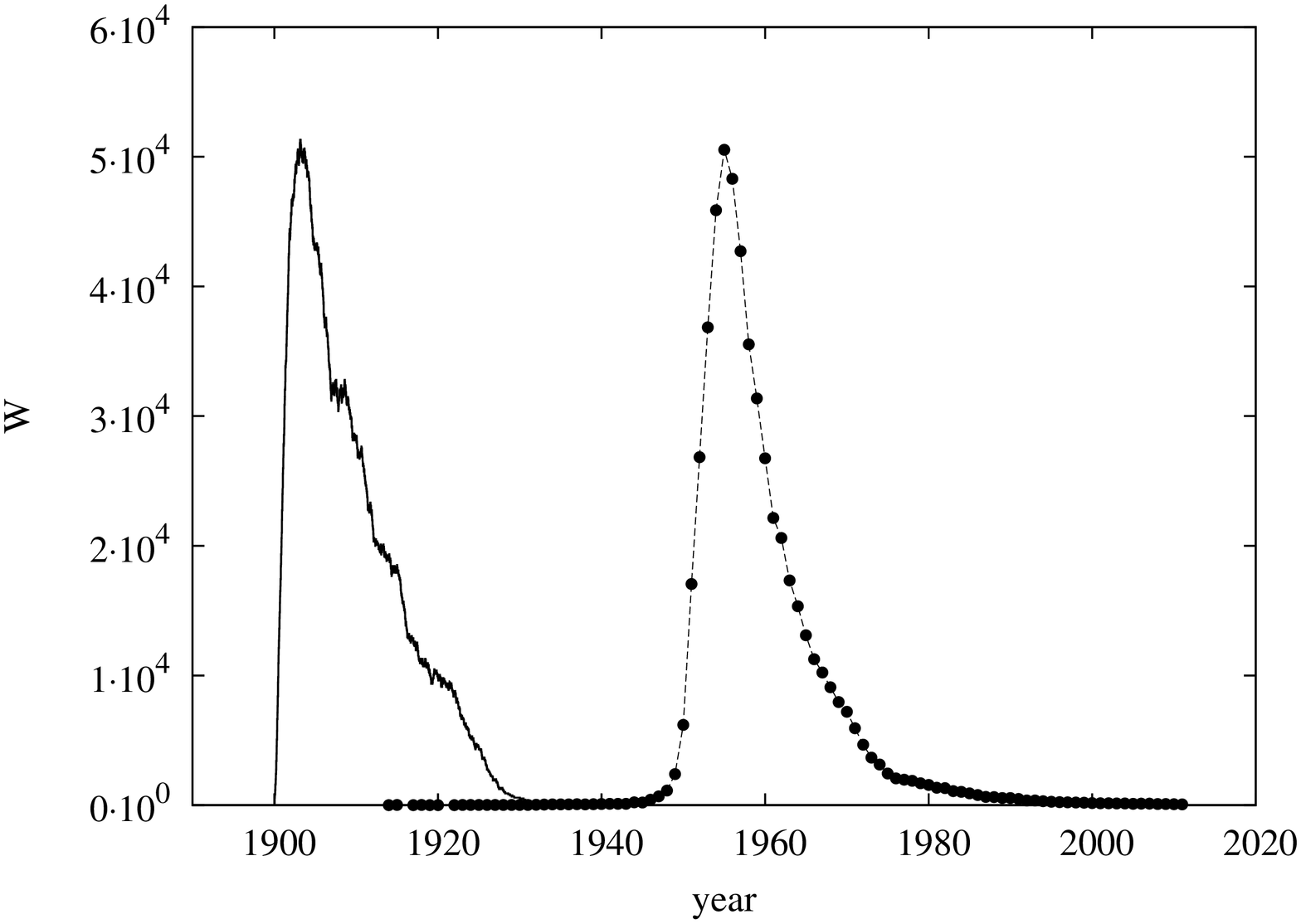}}}}
\caption{Time dependence of number $W$ of newborn children named Debra (circles) \cite{names} and a typical shape of a calculated plot.  In the latter,
the scales of both axes are adjusted.}
 \label{fig-6}
 \end{figure}

On the contrary to this seemingly universal behaviour, for small $q$ the system behaviour is different. Namely, for $q$=5
the observed values never disappear: once a value is present, it will be present forever, and the fluctuations of its
frequency remain relatively small. A typical example is shown in Fig. 2. As $q$ increases, a tendency towards smaller 
frequencies can be seen, and it prevails above $q$=20, where more and more symbols disappear. However, the change is not 
sharp, and the term "crossover" seems to be more appropriate than "transition". \\

We made an attempt to check if the time period $T$ of a typical "rise and fall" depends on $q$. Indeed, for $q=30$ this time 
is found to be about $71.4N$ time steps, while for $q=1000$ it is about $17.4N$ time steps, i.e. more than 4 times smaller. Here
 the calculation was made for $N=5000$. (The time step is equivalent to a check of a pair of nodes.) However, we note that the 
distribution of $T$ is rather wide; the statistics (413 cases for $q=30$, 2251 in the same time length of the sample for $q=10^3$)
does not allow to infer about the character of the plots, shown in Fig. 3.  Yet we can see that the cycle length decreases 
with $q$: more names, shorter the cycles.

\section{Data sets}

At the webpage of the U. S. Social Security Administration \cite{names}, the statistics is available on the frequency of 
male and female given names in 1880-2011. The datasets are provided also for particular states, but the latter is not analyzed
here. From these data, we extracted the time dependences of number of events, when a child got a given name in a given year.
These data show irregularities, which cannot be atributed to random noise. Actually, we found it fascinating to trace how the 
names of different kinds of celebrities are visible or not, appear and disappear over the course of the years. Sometimes the 
relation to a given person seems obvious, as for the name Charlie. The number of newborn boys with this name was almost constant
before 1910, but the data show an increase, more or less linear,
from about 800 in 1909 to almost 2900 in 1919. Then, the plot started to decrease so slowly that the plateau about 500 was obtained 
not earlier than in 1970. In other cases, an interpretation is less straightforward. In 1910, a similar increase of popularity 
is observed for the name Albert, from about 2000 to more than 10000 in 1921. In 1910, Einstein was proposed for the first time 
as a candidate for the Nobel Prize. However, some contribution is possible also from Albert I of Belgium, who started his reign
in 1909. A study on history of American culture could bring some light in this matter. To end with more recent example, a sharp increase
of number of small Angelinas from 1000 to almost 6000 between 2000 and 2005 does not leave doubts about its origin. \\

Two facts about the data \cite{names} can be of interest for further research. First, the diversity of names tends to increase in time.
This can be seen, for example, in the time dependence of the fragmentation index, defined as $S(t)=\sum_i(p_i(t))^2$, where $i$ 
is the name index and $p_i(t)$ is the percentage of babies with this name in $t$-th year. The plot $S(t)$ is shown in Fig. 4.
It is remarkable that despite the overall tendency of decrease, a small maximum is observed in 1947: again a question for cultural 
studies. The second fact is that the distribution of $p_i$ is close to the power law, and the quality of the fit does not worsen 
in time. In 2011, the exponent was 1.9 (see Fig. 5) and its older values are not far from this. We add that in our fitting procedure the first 
point is purposefully omitted. This point refers to most frequent names. Why it does not fit to the power law? this question
 should be discussed together with the question about the origin of the power law itself. \\

When compared with babies names, the data on skirt lengths are much less complete. In \cite{curran}, we get the mean skirt length ratios to
height of figure, in the UK and West Germany. The data come from measurements of photos of day dress in two autumn magazines, Littlewoods Ltd
and Neckermann Versand, in 1954-1990. Plots and tables are given also on the ratios of skirt width to height and on the standard deviations of skirt lengths and widths within the year in the same period. While the data on skirt lengths show approximately the same time dependence in both magazines, the 
ratio of width to height differ more clearly before 1970. We note that the standard deviation for skirt length is the largest after 1980,
while the one for skirt width is the largest before 1962. On the other hand, conclusions of an analysis of data from 1789-1980 \cite{low} 
suggest that according to the general trend, the within-year variance increase in time. \\

Earlier data reveal that the skirt length loses its discriminatory power in terms of today. In \cite{kro}, we find data from fashion journals for 1845-1915, where the accuracy is one milimeter. As we read there, the ratio of length of dress to height of figure was not smaller than 0.95 till 1912, with a sudden fall to 0.842 in 1919 - what a come down! In the same period the skirt diameter passed a variation from 55 cm to 110 cm, then to 30 cm. On the other hand, as we read in \cite{barbie} (the data from 1860 to 1980, four fashion journals), more shirt lengths were concurrently present and the measurements give only most typical results. In \cite{barbie}, the data are provided in a coarse-grained form; for skirt length, the categories are as follows: train, floor, ankle, calf, cover knee, above knee. The last category was rather broad already in 1980.\\

\section{The Simmel effect in names}

There are five 'classic' names, with their popularity top above 60 thousands babies born around 1946-1950,
with remarkable maxima also at 1920 or a few years later. These are: James, John, Mary, Robert and William.
In the case of Mary, the top at 1920 is the highest. However, the top record belongs to Linda (see Fig.6). 
This name, popular already in 1946 (52 thousands), jumped to above 99 thousands in 1947; we believe that the effect
is due to a popular song entitled 'Linda', released in November 1946. Other names from 'top ten' are: Michael, 
David and Jennifer with broad maxima in 1957, 1960 and 1972, respectively, and Lisa: a sharper maximum in 1965.\\

In the next ten names, again we find peaks . In five of them, the slope is larger on the left, as in 'lambda'.
Four of them is more or less symmetric, and in only one (Patricia) perhaps the slope is larger on the right. 
 We stress that after inspecting a hundred of most popular names, we 
have always seen a structure which cannot be attributed to uncorrelated fluctuations. In most cases, an eye 
armed with the Rayleigh criterion sees only one maximum. Often, the lambda shape can be identified: an example 
is given in Fig. 7, together with a calculated plot which can be considered as 'typical'.  
From numerous associations to contemporary celebrities, we mention the shortest career of Shirley, from 14316 newborn babies
in 1933 to 35149 in 1936, but less than 18 thousands already in 1940.\\

\section{Discussion}

Our main goal is to connect the approach of Pedone and Conte \cite{con1,con2} with the reference data on fashion. Indeed, 
the model captures the difference between the case of small $q$ (say, 5) and large $q$ (30 or more). In the former case, particular values of $x$ never disappear; in the latter, 
clear cycles are observed, cf. Fig. 1 vs Fig.2. This follows the difference between the length of skirts \cite{curran,barbie,low} and 
the babies names \cite{names}. \\

A critical comment is that the skirt length is a continuous variable and the classification in literature (above ankle, calf etc.)
is arbitrarily rough. If we enter into details of a given female dress, we could find that the pattern used, for example, by two 
ladies in Natchez in 1848 was never repeated in later history.  On the other hand, if the length of names is used instead of 
name itself, the results on name sets could be the same as those for the skirt length. Then, the above difference can be assigned 
not as to the difference between the data sets but rather to our rule of classification. It is even possible to draw an analogy of 
the phenomenon of fashion with the idea of micro- and macroscopic states. A continuation of his thread is, however, out of the scope 
of this text. \\

We conclude that the combination of imitation and avoidance allows to interpret the dynamics of frequency of babies names. 
In this interpretation, the rise and fall of popularity of a name is a consequence of social imitation and avoidance. We note that
today, these changes of popularity are more abrupt that in times of Simmel. Then, the correspondence of his theory with the contemporary data 
was not as clear yet as it is today.

 \section*{Acknowledgements} The paper is dedicated to Dietrich Stauffer on the occasion of his future 70-th birthday.

\end{document}